\documentclass[12pt]{article}
\pdfoutput=1
\usepackage{hyperref}
\usepackage{color}
\usepackage{graphicx}
\usepackage{amsmath}
\usepackage{amssymb}
\usepackage{xspace}
\usepackage{soul}

\makeatletter
\@addtoreset{equation}{section}

\makeatletter
\renewcommand\section{\@startsection {section}{1}{\z@}%
                                   {-3.5ex \@plus -1ex \@minus -.2ex}%nn
                                   {2.3ex \@plus.2ex}%
                                   {\normalfont\large\bfseries}}
\renewcommand\subsection{\@startsection{subsection}{2}{\z@}%
                                     {-3.25ex\@plus -1ex \@minus -.2ex}%
                                     {1.5ex \@plus .2ex}%
                                     {\normalfont\bfseries}}

\newcommand{\be}{\begin{equation}}
\newcommand{\ee}{\end{equation}}
\newcommand{\beq}{\begin{eqnarray}}
\newcommand{\eeq}{\end{eqnarray}}

\newcommand{\gone}[1]{{}}

\begin{document}
\begin{titlepage}

%\hfill Last compiled \today \ at \currenttime\\  
\rule{0ex}{0ex}

\vfil

\begin{center}

{\bf \Large
  Thermodynamics of $T \bar T$, $J \bar T$, $T \bar J$ deformed \\
  conformal field theories
}

\vfil

Soumangsu Chakraborty$^1$,  Akikazu Hashimoto$^2$

\vfil

{}$^1$ Department of Theoretical Physics,\\
Tata Institute for Fundamental Research, Mumbai 400005, India

{}$^2$ Department of Physics, University of Wisconsin, Madison, WI 53706, USA

\vfil

\end{center}

\begin{abstract}
We compute the Hagedorn temperature of $\mu T \bar T + \varepsilon_+ J \bar T + \varepsilon_-T \bar J$ deformed CFT using the universal kernel formula for the thermal partition function. We find a closed analytic expression for the free energy and the Hagedorn temperature as a function of $\mu$, $\varepsilon_+$, and $\varepsilon_-$ for the case of a compact scalar boson by taking the large volume limit.  We also compute the Hagedorn temperature for the single trace deformed $AdS_3 \times S^1 \times T^3  \times S^3$ using holographic methods. We identify  black hole configurations whose thermodynamics matches the functional dependence on $(\mu, \varepsilon_+, \varepsilon_-)$ of the double trace deformed compact scalars.

\end{abstract}
\vspace{0.5in}

\end{titlepage}

\section{Introduction}

Recently, $\mu T \bar T$ and related deformation of quantum field theories in 1+1 dimensions have been a subject of great interest. Here, $T \bar T$ is a composite operator built in terms of the stress tensor, and $\mu$ is its coefficient whose dimension is square of length. Such a deformation is non-renormalizable. Despite that, it was demonstrated in \cite{Smirnov:2016lqw} that the spectrum of deformed theory living on a cylinder can be determined if the spectrum of the undeformed theory is given. After the deformation, the system appears to exhibit features of non-locality. One clear indication of this is the Hagedorn spectrum in the UV. $T \bar T$ deformation therefore appears to be a way to extend the usual notion of quantum field theories in a controlled setting. 

A robust observable one can compute for these class of theories is the thermal partition function as a function of inverse temperature $\beta$ and the radius $R$. Since the spectrum is known, the partition function is determined unambiguously. One can also think of this observable as a vacuum amplitude for the Euclidean theory living on a torus whose periods are $\beta$ and $2 \pi R$. Through parameters $\beta$ and $R$, one can probe the various scales of the system. To keep the discussion simple, it is convenient to restrict our attention to the case where the undeformed theory is a conformal field theory. Then, the deformation introduces a single scale $\mu$, which we can probe using $R$ and $\beta$.

The torus partition function have many interesting properties and have been computed by many authors. First, a flow equation for the partition function was derived in \cite{Cardy:2018sdv}, and a solution to this flow equation was computed using the JT gravity formulation of $T \bar T$ deformation in \cite{Dubovsky:2018bmo}. In a parallel development, authors of \cite{Aharony:2018bad,Datta:2018thy} highlighted the power of modular properties in computing the partition function.

In a theory with $U(1)$ global symmetry with holomorphic and anti-holomorphic currents $J$ and $\bar{J}$ respectively, there are also  deformations by $\varepsilon_+ J \bar T$   and $\varepsilon_- T \bar J$ which also turn out to be integrable \cite{Guica:2017lia,Chakraborty:2018vja}. The partition function also exhibits modular property \cite{Aharony:2018ics}.

These modular properties are suggestive of the connection between $T \bar T$ and related deformations and the world-sheet sigma model of string theory.\footnote{The connection between $T \bar T$ deformation and Nambu-Goto action was first derived in \cite{Cavaglia:2016oda}. See also \cite{Bonelli:2018kik}.} In order to utilize this relation, the authors of \cite{Chakraborty:2019mdf} integrated the deformation of the world-sheet sigma model by an exactly marginal world-sheet operator whose space-time interpretation is the deformation by irrelevant operators whose coefficients are $(\mu,\varepsilon_+,\varepsilon_-)$. Strictly speaking, the system being considered was that of a $1+1$  dimensional  CFT realized holographically as $AdS_3 \times {\cal M}$ in type IIB string theory, deformed by an integrable single trace $T \bar T$ deformation \cite{Giveon:2017nie}. However,  the spectrum of states on the long strings sitting in the weakly coupled region can be interpreted as experiencing the double trace  deformation  $\mu T \bar T + \varepsilon_+ J \bar T + \varepsilon_- T \bar J$. In this way, and by invoking universality, the authors \cite{Chakraborty:2019mdf} were able to compute the spectrum of states for the world-sheet sigma model whose target space is some generic ${\cal M} = S^1 \times {\cal M}'$. The resulting spectrum is in agreement with \cite{LeFloch:2019rut,Frolov:2019xzi} who followed the approach of \cite{Smirnov:2016lqw,Guica:2017lia} more closely.

More recently, a direct computation of the world-sheet torus amplitude for the sigma model on $T^2 \times S^1 \times {\cal M}'$, restricted to the sector where the world-sheet wraps the $T^2$ exactly once, was presented as a compact integral kernel (\ref{Zdef})  acting on the partition function of the undeformed theory in \cite{Hashimoto:2019wct}. This expression made the modular properties manifest. It also reproduced the spectrum obtained previously in  \cite{Chakraborty:2019mdf,LeFloch:2019rut,Frolov:2019xzi}.

In this article, we will explore the application of kernel formula, derived in \cite{Hashimoto:2019wct}, on  a single free compact boson CFT. More concretely, we will compute how the Hagedorn temperature depends on $\mu$, $\varepsilon_+$, and $\varepsilon_-$.  This analysis enables us to identify points which are special in the $(\mu, \varepsilon_+, \varepsilon_-)$ parameter space. 

To facilitate the analysis, we found it convenient to analyze the thermodynamics in the large volume limit $R \rightarrow \infty$ keeping $\beta$, $\mu$, $\varepsilon_+$, and $\varepsilon_-$ fixed. Although the thermodynamic quantities do simplify dramatically, some critical information also gets lost in the limit.  We will elaborate on the implication of this issue in the following sections.   We will also compute the Hagedorn temperature of the holographically realized single trace deformed systems \cite{Giveon:2017nie,Chakraborty:2018vja,Chakraborty:2019mdf} and comment on their features. 

\section{Review of the kernel formula and universality}

In this section, we will review the master formula for computing the partition function of generic $T \bar T$, $J \bar T$, $T \bar J$ deformed CFT derived in \cite{Hashimoto:2019wct}. It is stated compactly as 
\be Z_{def}(\zeta, \bar \zeta, \lambda, \epsilon_+, \epsilon_-)= \int_{{\cal H}} d^2 \tau \int_{{\cal C}} d \chi d \bar \chi \ I(\zeta, \bar \zeta, h, \epsilon_+ \epsilon_-, \tau, \bar \tau, \chi, \bar \chi) Z_{inv}(\tau, \bar \tau, \chi, \bar \chi)  \label{Zdef}~,  \ee
with
\be I = {1 \over 4 \epsilon_+ \epsilon_- \tau_2^3} \exp\left[
-{\pi \chi \bar \chi \over2 h\epsilon_+ \epsilon_- \tau_2} 
  - {\pi \chi \over 2 \epsilon_+ \tau_2} (\bar \tau - \bar \zeta)
 + {\pi \bar \chi \over 2 \epsilon_-  \tau_2} (\tau - \zeta)
\right]\ ,  \label{I}\ee
where we introduced dimensionless deformation parameters
\be \lambda = {\mu \over R^2} , \qquad  \epsilon_+ = {\varepsilon_+ \over R}, \qquad \epsilon_- = {\varepsilon_- \over R}\  .  \ee
The quantity $Z_{inv}$ is defined as
\be Z_{inv}(\chi, \bar \chi) = e^{\kappa\pi (\chi - \bar \chi)^2/2 \tau_2}  Z_{cft}(\chi, \bar \chi) \label{Zinv}\ee
where  $\kappa$ is the level of the $U(1)$ current which we set to 1 for the case of single compact scalar which  we will consider as our main example, and
\be Z_{cft}(\tau,\bar \tau, \chi, \bar \chi) = \sum_i e^{- 2 \pi \tau_2 E_i + 2 \pi i \tau_1 p_i + 2 \pi i \chi p_{Li}  - 2 \pi i \bar \chi p_{Ri}  }  \label{Boltzmann}\ee
is the Boltzmann sum for a CFT with charges weighted by chemical potential parameters $\chi$ and $\bar \chi$.  These formulas are taken from (4.10), (4.11), (A.7), and (A.13) of  \cite{Hashimoto:2019wct}. The formula (\ref{Zdef})  was derived by manipulating the sigma model for long strings in the single trace deformed theory of \cite{Chakraborty:2019mdf} where the $U(1)$ isometry was some compact $U(1)$. Since the effects of $T \bar T$, $J \bar T$, and $T \bar J$ deformation acts universally, we can regard (\ref{Zdef}), (\ref{I}), and (\ref{Zinv}) to also be universal. The fact that $Z_{def}$ is invariant under
\be \zeta \rightarrow {a \zeta + b \over c \zeta + d}, \qquad \lambda \rightarrow {\lambda \over |c \zeta + d|^2}, \qquad \epsilon_+ \rightarrow {\epsilon_+ \over c \bar \zeta + d}, \qquad \epsilon_- \rightarrow {\epsilon_- \over c \zeta + d} \label{zetamod} \ee
was also explained in  \cite{Hashimoto:2019wct}.

The parameters $h$ and $\lambda$ are related according to
\be h^{-1} = \lambda + a \epsilon_+ \epsilon_- \ee
where $a$ is some constant which contributes to $I$ as a factor of $e^{- \pi a \chi \bar \chi / 2 \tau_2}$ which can be absorbed into $Z_{inv}$ without affecting the modular properties. This freedom is a manifestation of changing the contact term between two $U(1)$ current operators explained in footnote 2 of \cite{Hashimoto:2019wct} and in  \cite{Kutasov:1988xb}. 

The goal of this article is to apply the formulas  (\ref{Zdef}), (\ref{I}), and (\ref{Zinv}) to extract some physical features. Specifically, we will compute the Hagedorn temperature and the free energy for CFT for a free compact scalar field theory. These quantities simplifies significantly in the infinite volume limit and can be presented in closed form. We will comment on physical features which can be inferred from these results.

\section{Thermodynamics and the infinite volume limit}

 In this section, we will apply the universal formula (\ref{Zdef}) and (\ref{I}) to analyze the thermodynamics of the deformed theory. The dimensionful parameters are deformation parameters $(\mu, \varepsilon_+, \varepsilon_-)$, inverse temperature $\beta$, and radius $R$. We can probe the $(\mu, \varepsilon_+, \varepsilon_-)$ dependence of the free energy and related thermodynamic quantities as a function of the temperature. From this point of view, the finite size effect when $\beta \sim R$ is not interesting. We can therefore scale $R$ out of the problem by sending it to infinity, keeping other dimensionful parameters finite.  This will substantially simplify the expression for the free energy and related quantities.\footnote{Thermodyanmics of $T \bar T$ deformed CFT at next to leading order in the large volume limit can be found in  \cite{Barbon:2020amo}.}
 
 We also need to address the dependence on $\zeta_1$ which is related to momentum along the spatial circle. It turns out that a convenient choice is to integrate $\zeta_1$ in the range $0 \le \zeta_1 \le 1$ which amounts to considering the thermodynamics in the zero momentum sector
 
\be Z^0_{def}(\zeta_2) = \int_0^1 d \zeta_1 \ Z_{def}(\zeta, \bar \zeta) \ . \ee
One can then show that
\be Z^0_{def}(\zeta_2, \lambda, \epsilon_+, \epsilon_-) = \int_{{\cal H}} d^2 \tau \int_{{\cal C}} d \chi d \bar \chi \ I(\zeta, \bar \zeta, \lambda, \epsilon_+ \epsilon_-, \tau, \bar \tau, \chi, \bar \chi) Z^0_{inv}(\tau_2, \chi, \bar \chi) ~,\label{Zno1}\ee
where
\be
Z^0_{inv}(\tau_2, \chi, \bar \chi)  = \int_0^1 d \tau_1 \ Z_{inv}(\tau, \bar \tau, \chi, \bar \chi) \ .
\ee
Even though the right hand side of (\ref{Zno1}) appears to depend on $\zeta_1$, it only appears in the combination $\tau_1 - \zeta_1$, and this is the only dependence on $\tau_1$ and $\zeta_1$.  Therefore the  dependence on $\zeta_1$ in (\ref{Zno1}) disappears upon integrating over $\tau_1$.

 Let us now describe how one extracts the large $R$ limit. When $\epsilon_\pm=0$, 
\be h = \lambda^{-1} = {R^2 \over \mu} \ee 
and so the large $R$ limit corresponds to large $h$. We can isolate the large $h$ scaling behavior by introducing the rescaling
\begin{eqnarray}
 \tau_i &=& {1 \over \sqrt{h}} t_i ~,\nonumber \\
 \zeta_i & = & {1 \over \sqrt{h}} z_i ~,\\
 \epsilon_\pm & = & {e_\pm \over \sqrt{h}} ~,\nonumber
 \end{eqnarray}
 so that
\be
  I \sim \exp\left[ -{\pi \sqrt{h} ((t_1 - z_1)^2 + (t_2 - z_2)^2) \over 2 t_2} - 
  {\pi \sqrt{h}\over 2 (e_+ e_-) t_2}(\bar \chi +e_- (\bar t- \bar z))(\chi - e_+ (t - z))\right] \ . \ee
We see then that in the large $h$ limit, the kernel is localized so that one can use the saddle point approximation. If one is only interested in the leading large $h$ behavior, we can also ignore the factor outside the exponential in $I$. We also see that the saddle point is dominated at small values of $\tau_2$ of order $h^{-1/2}$.  Working in the regime,
\be h^{-1} \sim \epsilon_+^2 \sim \epsilon_-^2 \ll 1, \ee
is therefore a natural scaling limit to explore the behavior of Hagedorn temperature on these variables. 

\subsection{Hagedorn temperature for the pure $T \bar T$ deformation.}

Let us begin by analyzing the simple case where $\epsilon_+$ and $\epsilon_-$ is set to zero. Then, the $\chi$ integral is localized at $\chi=0$ so $Z_{inv}=Z_{cft}$. We can take a generic CFT with central charge $c$ as the undeformed theory. This is a trivial case where the answer is known from previous works, but it provides a concrete template which we can use as a guide in considering more complicated cases.

 The partition function in the small $\tau$ limit is given by the Cardy's formula
\be Z_{cft}[\tau, \bar \tau] = \exp\left[ {i\pi c \over 12 \tau} - {i\pi c \over 12 \bar \tau}\right] \ . \ee
We need to integrate over $\tau_1$ in order to isolate the zero momentum sector. Fortunately, in our scaling limit, this integral is localized at $\tau_1=0$. So we can consider 
\be Z^0_{cft} = Z^0_{inv} = \exp\left[{\pi c \over 6 \tau_2} \right]  \ee
to be our starting point. 
The remaining $\tau_1$ integral is a trivial Gaussian integral. All that remains then is to find the saddle point for $\tau_2$ for the action
\be   -{\pi h (\tau_2 - \zeta_2)^2 \over 2 \tau_2} + {\pi c \over 6 \tau_2} \ . \ee
This leads to 
\be \log Z_{def}^0(\zeta_2) = {\pi c \over 3 \zeta_H^2} \left( \zeta_2 - \sqrt{\zeta_2^2 - \zeta_H^2}\right)~, \ee
where 
\be \zeta_H = \sqrt{{c \over 3 h}} \ee
is the branch point in the $\zeta_2$ dependence of the partition function. This expression contains all the information about the thermodynamic potentials and the equation of state up to standard Maxwell relations.

It is straight forward to Legendre transform the partition function to obtain the equation of state (thermal entropy)
\be S = 2 \pi  \sqrt{  {c R {\cal E} \over 3} + \zeta_H^2  R^2 {\cal E}^2} ~,\ee
from which we read off the Hagedorn temperature
\be \beta_H = 2\pi R \zeta_H = 2 \pi \sqrt{c \over 3} \sqrt{R^2 \over h} = 2 \pi \sqrt{c \mu \over 3}~, \label{betaHtt} \ee
which is finite in the scaling limit.

A useful comparison is to look at the energy of the lowest energy state, 
\be R {\cal E}_0 = \sqrt{{1 \over 4 \lambda^2} + {R E_0 \over \lambda} + (R P_0)^2 } -{1 \over 2 \lambda},\ee
which for
\be R E_0 = -{c \over 12}, \qquad R P_0 = 0 \ , \ee
leads to a branching behavior at
\be {1 \over \lambda} = {c \over 3} \ee
or 
\be2 \pi  R = 2 \pi \sqrt{c\mu  \over 3} .\ee
This is a reflection of the fact that infinite volume at finite Euclidian time coordinate with periodicity $\beta$ is geometrically equivalent to finite volume of period $2 \pi R$ with infinitely extended time coordinate. The Hagedorn behavior from the first point of view corresponds to the appearance of tachyon from the second point of view. This point was also emphasized in  \cite{Aharony:2018bad} and will continue to hold for the more general cases we will be considering below.

\subsection{Compact Scalars \label{sec:kernel}}

Now that the template for analyzing the thermodynamics of pure $T \bar T$ deformation is established, it is straightforward to extend the analysis to the case of  compact  scalars.

Let us consider the case of a single compact scalar. What we need is the large $R$ limit of $Z_{inv}$. We know that $Z_{cft}$ for this case is 
\be Z_{cft}(\tau,\bar \tau, \chi, \bar \chi)=  |\eta(\tau)|^{-2} Z_{\rm zeromode} ~,\ee
where, using (A.9) of  \cite{Hashimoto:2019wct}, we have
\be Z_{\rm zero mode} =  \sum_{n,w} \exp \left[-2 \pi \tau_2 \left({n^2 \over r^2} + {w^2 r^2 \over 4} \right) + 2 \pi i \tau_1 n w+
  2 \pi i \chi \left({n \over r} + {w r \over 2}\right)
   - 2 \pi i \bar \chi \left({n \over r} - {w r \over 2}\right) \right] \ . \label{charges} \ee
Taking the infinite volume limit is equivalent to taking $\tau_2$ to be small. In this limit, the sum over $n$ and $w$ can be approximated by an integral, and we find
\be Z_{cft}(\tau, \bar \tau,  \chi, \bar \chi) = {1 \over \tau_2^2} e^{{\pi \over \tau_2} \left( {1 \over 6} - \chi^2 - \bar \chi^2 \right)} \label{Zchi}\ee
and
\be Z_{inv}(\tau, \bar \tau, \chi, \bar \chi) =\exp\left[ {\pi (\chi - \bar \chi)^2 \over 2 \tau_2} \right] Z_{cft} (\tau, \bar \tau, \chi, \bar \chi) \ . \ee
All that remains to be done is to integrate out $\chi$, $\bar \chi$, $\tau_1$, and $\tau_2$ of (\ref{Zdef}) in saddle point approximation.  This only requires some algebraic manipulations, and we find 
\be \log Z_{def}^0(\zeta_2,\epsilon_+, \epsilon_-) = {\pi  \over 3 \zeta_H^2} \left( \zeta_2 - \sqrt{\zeta_2^2 - \zeta_H^2}\right)  ~,\label{ZzH}\ee
where 
\be \zeta_H = \sqrt{{\lambda \over 3} + {1 \over 3} (\epsilon_+ - \epsilon_-)^2}   \ee
and we are setting $h^{-1} = \lambda - 4 \epsilon_+ \epsilon_-$ following  \cite{Chakraborty:2019mdf,Hashimoto:2019wct}.

The functional form of the partition function (\ref{ZzH})  in terms of $\zeta_2$ and $\zeta_H$ is identical to the pure $T \bar T$ deformation. This was somewhat unexpected. The dependence on $\epsilon_+$ and $\epsilon_-$ only appear in the combination which enters in $\zeta_H$.  The conclusion then is that the inverse Hagedorn temperature is given by 
\be \beta_H = 2 \pi  \sqrt{{1 \over 3} \left(\mu  +(\varepsilon_+ - \varepsilon_-)^2\right)} \ . \label{result1}\ee

One can further verify the validity of (\ref{result1}) by exchanging the $x_1 \leftrightarrow(-x_2)$ flip and looking at the mass of the ground state. This exchange of $x_1$ and $(-x_2)$ is essentially the  modular transformation $\zeta \rightarrow -1/\zeta$. Assuming that $\zeta=i$ is purely imaginary,  They will transform as
%i
\be \epsilon_+ \leftrightarrow  i \epsilon_+~, \qquad \epsilon_- \leftrightarrow -i \epsilon_-  \ . \label{exchange}\ee

The energy of the ground state can be read off from (4.14) of \cite{Hashimoto:2019wct} and is
\be ER = {1 \over 2 A} \left(-B - \sqrt{B^2 - 4 A C}\right), \label{ERdef}\ee
with
\beq
A & = & - {1 \over R^2}(\mu - (\varepsilon_+ + \varepsilon_-)^2), \cr
B & = & -1, \cr
C & = & -{c \over 12}, \label{ABC} \eeq
which has a branch point when
\be  2 \pi R =  \sqrt{ {c \over 3}(\mu  - (\varepsilon_+ + \varepsilon_-)^2)} \ee
where, despite the  appearance,  the right hand side is independent of $R$, and agrees with (\ref{result1}) upon mapping the transformation (\ref{exchange}). Since the Hagedorn behavior should correspond to appearance of tachyon in $x_1 \leftrightarrow (- x_2)$ flip, this is a non-trivial check on the validity of (\ref{result1}).

One might wonder which states are contributing to the Hagedorn density (\ref{result1}). One way to address this question is to explore the spectrum associated with individual $(n,w)$ sectors of (\ref{charges}) separately.  The flow equation (4.13) of  \cite{Hashimoto:2019wct} for zero momentum relates the undeformed and deformed energy by
\be RE = (1 + 2 \epsilon_+ q_L - 2 \epsilon_- q_R) R {\cal E}  + (\lambda - (\epsilon_+ + \epsilon_-)^2) R^2 {\cal E}^2, \ee
where
\be q_L = {n \over r} + {w r \over 2} , \qquad q_R = {n \over r} - {w r \over 2} \ . \ee
The undeformed CFT in fixed $(q_L,q_R)$  charge sector has entropy
\be S^{q_L,q_R} (E) = 2 \pi \sqrt{ {1 \over 3}R E - {1 \over 6} (q_L^2 + q_R^2)} \ . \ee
One can therefore write
\be S^{q_L,q_R}({\cal E}) = 2 \pi \sqrt{ {1 \over 3}\left( \rule{0ex}{2.5ex}(1 + 2 \epsilon_+ q_L - 2 \epsilon_- q_R) R {\cal E}  + (\lambda - (\epsilon_+ + \epsilon_-)^2) R^2 {\cal E}^2 \right)  - {1 \over 6} (q_L^2 + q_R^2)} \  .\label{entmc} \ee
At energy ${\cal E}$, the  dominant contribution of the entropy comes from the sector
\be q_L = 2 \varepsilon_+ {\cal E}, \qquad q_R = - 2 \varepsilon_- {\cal E} \label{domcharge}\ee
for which 
\be S^{q_L,q_R}({\cal E}) = 2 \pi \sqrt{{1 \over 3} \left( R{\cal E} + (\mu + (\varepsilon_+ - \varepsilon_-)^2) {\cal E}^2\right) } \label{SqLqR}\ee
and we can read off the Hagedorn density  (\ref{result1}) from the coefficient of the ${\cal E}^2$ term inside the square root.

It is interesting to note that for fixed $(q_L,q_R)$, the Hagedorn density read off from (\ref{entmc}) 
\be \beta_H^{q_L,q_R} = 2 \pi \sqrt{{1 \over 3} (\mu - (\varepsilon_+ +\varepsilon_-)^2)}   \label{betaA} \ee
is a different quantity than (\ref{result1}). One way to describe the situation is that the grand canonical and fixed charge ensemble leads to different Hagedorn densities.  It is also notable that if condition
\be - A =  \lambda - (\epsilon_+ + \epsilon_-)^2 > 0  \label{Acond} \ee
is not satisfied, fixed charge ensembles are ill defined. The quantity $A$ appeared previously in (\ref{ABC}) and indicates that energy of infinitely many states are becoming complex.   This can also be seen from the fact that $ \beta_H^{q_L,q_R}$ is not a real when (\ref{Acond}) is not satisfied. A closely related fact noted in \cite{Hashimoto:2019wct} is the fact that integral over $\tau_2$ in the kernel formula (\ref{Zdef}) is unbounded when (\ref{Acond}) is not satisfied.  The grand canonical ensemble must also be ill defined since it is a sum over charge sectors. Even though (\ref{ZzH}) does not show any pathology when $(-A)$ flip sign, it and (\ref{result1}) should be considered valid only when (\ref{Acond}) is satisfied.

\section{Thermodynamics of holographic single trace deformed system \label{sec:sugra}}

In this section, we will examine the thermodynamics of  single trace $T \bar T$, $J \bar T$, $T \bar J$ deformed CFT constructed holographically. The prototype of this construction is \cite{Giveon:2017nie,Israel:2003ry} where a background corresponding to a scaling limit of NS5-F1 system  was considered. This geometry behaves in core region as $AdS_3\times T^4 \times S^3$, whereas in  the asymptotic region it asymptotes to  a linear dilaton geometry $R^2 \times R^\phi \times T^4 \times S^3$.  In \cite{Giveon:2017nie}, this background was derived by integrating the deformation of world-sheet sigma model by an operator of the form
\be  \lambda \int d^2 \sigma \  J^-\bar{J}^-  ,  \ee
where $J^-$ and $\bar{J}^-$ are respectively the left and right-moving null $SL(2,\mathbb{R})$ currents. Here, $\lambda$ is the coefficient of the deformation operator of the world-sheet sigma model. The point is that this deformation corresponds to the deformation of the target space theory which is a holographic dual of a deformed CFT.  So we start with $AdS_3 \times S^3 \times T^4$ constructed by taking the near horizon limit of a stack of NS5 and F1.  This background in the supergravity language can be written in the form
\be {ds^2 \over \alpha'} = h d \gamma d \bar \gamma +  d \phi^2 + dy^2 + ds^2_{T^3} + k  ds^2_{S^3} ~,\label{ttbackground} \ee
where we are working in Minkowski signature
\be \gamma = \gamma_1 + \gamma_0, \qquad \bar \gamma = \gamma_1 - \gamma_0 \ . \ee
 We have also identified one of the coordinates of $T^4$ as $y$.

A comment is in order that $h(\phi)$ 
\be h(\phi)^{-1} = {\alpha' \over R^2} + e^{-2 \phi} = \lambda + e^{-2 \phi} \  \label{hrel} \ee
in this context is a field with non-trivial profile along the $\phi$ direction. This is in contrast to the fact that $h$ was treated as a parameter in the previous sections. One way to think about this is the fact that the sigma model treatment in the earlier section was for a long string sitting in the weakly coupled region as was the case in \cite{Chakraborty:2019mdf,Hashimoto:2019wct}. There are other fields such as the dilaton and the form fields which we omit here for brevity but can be found in  \cite{Giveon:2017nie}. For the pure single trace $T \bar T$ deformation,  the deformation parameter $\lambda$ turns out to be equal to $\alpha'/R^2$.   

A useful observation to make at this point is that the string theory background found in \cite{Chakraborty:2019mdf} can be reconstructed starting with (\ref{ttbackground}) and performing the following operations:
\begin{enumerate}
\item Twist the $(\gamma, \bar \gamma, y)$ coordinates by
\be \gamma \rightarrow \gamma + 2 \epsilon_+ y, \qquad \bar \gamma \rightarrow \bar \gamma+ 2 \epsilon_- y \label{twist1} \ee
\item Shift
\be h^{-1} \rightarrow h^{-1} - 4 \epsilon_+ \epsilon_- ~.\label{hshift} \ee
\end{enumerate}

Somewhat remarkably, one can reconstruct the same background starting with (\ref{ttbackground}) and acting with the following sequence of operations:
\begin{enumerate}
\item T-dualize from $y$ to $\tilde y$
\item Perform a twist
\be \gamma \rightarrow \gamma - 2 \epsilon_+ \tilde y, \qquad \bar \gamma \rightarrow \bar \gamma+ 2 \epsilon_- \tilde y \label{twist2} \ee
\item T-dualize back from $\tilde y$ to $y$,
\end{enumerate}
as well as
\begin{enumerate}
\item T-dualize on $\gamma_1$ and $\gamma_2$
\item Twist according to
\be y \rightarrow y - \epsilon_+ \tilde \gamma + \epsilon_- \bar {\tilde \gamma} , \label{epemtwist} \ee
\item T-dualize back to $\gamma_1$ and $\gamma_2$
\end{enumerate}
as was noted in \cite{Apolo:2018qpq,Apolo:2019yfj}.

In the end, one arrives at 
\be {ds^2 \over \alpha'} = {1 \over h(\phi)^{-1} - 4 \epsilon_+ \epsilon_-}  (d \gamma + 2 \epsilon_+ dy)(d \bar \gamma+ 2 \epsilon_- dy) +  d \phi^2 + dy^2 + ds^2_{T^3} + k ds^2_{S^3}~,  \label{ttbackground2} \ee
with $h(\phi)$ given by (\ref{hrel}).  Dimensionally reducing along $y$, $T^3$, and $S^3$ will bring this background to match the form presented in (4.8) of \cite{Chakraborty:2019mdf}.

Let us first recall the analysis of the thermodynamic behavior for this system with $\epsilon_+ = \epsilon_- = 0$. 
The finite temperature generalization of (\ref{ttbackground}) can be constructed from the non-extremal five dimensional black hole solution which can be read off from (2.44) of \cite{Maldacena:1996ky} up to S-duality and some convention mapping. The thermodynamics of this system can then be read off from the property of the horizon and the analysis of the conical singularity of the Euclidean solution. The intermediate steps of this analysis is somewhat cumbersome, but we will be brief here as the analysis is standard and it was essentially carried out in \cite{Giveon:2017nie,Apolo:2019zai}. The essential step is to solve for $r_0$ from  (2.50) of  \cite{Maldacena:1996ky}
\be T(r_0) = {1 \over 2 \pi r_0} \cosh \alpha \cosh \gamma \cosh \sigma \ee
and substitute into (2.49)  of  \cite{Maldacena:1996ky}
\be S(r_0) = {2 \pi R V r_0^3 \over g^2 \alpha'^4} \cosh \alpha \cosh \gamma \cosh \sigma \ee
subject to charges given in (2.45)  of  \cite{Maldacena:1996ky}
\be Q_1 = {V r_0^2 \over 2 g \alpha'^3 } \sinh 2 \alpha, \qquad Q_5 = {r_0^2 \over 2 g \alpha'} \sinh 2 \gamma, \qquad N = {R^2 V r_0^2 \over 2 g^2 \alpha'^4} \sinh 2 \sigma  \ . \ee
For our purposes we set  $N=0$ for convenience, and  $Q_5$ is taken to be asymptotically large to decouple the asymptotically flat region to make the spacetime asymptote to a linear dilaton geometry.  We have reinstated the factors of $\alpha'$ which was set to 1 in \cite{Maldacena:1996ky}. Applying S-duality to the resulting expression gives
\be S(T) \sim { Q_1  Q_5 R T \over \sqrt{1 -  \alpha' Q_5 T^2} } = { Q_1  Q_5 R T \over \sqrt{1 -  \lambda  Q_5 R^2 T^2} }~,\label{sugraST}\ee
where in the last equality, we used the fact that  $\alpha' = \lambda R^2$ from (\ref{hrel}). From  (\ref{sugraST}), 
we read off the expected Cardy behavior for small $T$,  and the Hagedorn behavior at
\be T_H \sim {1 \over Q_5^{1/2} \lambda^{1/2} R} ~. \label{sugraTH} \ee
In (\ref{sugraST}) and (\ref{sugraTH}),  $Q_1$ and $Q_5$ are respectively the F1 string and NS5 brane charges that make up the background. 
The $\sim$ is used to indicate that we have dropped   factors of order one such as $2$ and $\pi$. 
These results are in agreement with the results of \cite{Giveon:2017nie} and is qualitatively in agreement with what we found in (\ref{betaHtt}) up to factors of $c$ and $Q_5$. One of course does not expect exact agreement as double trace and single trace $T \bar T$ deformations are physically distinct \cite{Giveon:2017nie}.  

The next logical step is to include the effects of the $\epsilon_\pm$ deformation. The zero temperature and finite temperature solutions both have the same isometries along which we twist and T-dualize. So, the finite temperature solution can be generated readily by applying the same solution generating transformations,  similar in spirit to the solution generating technique utilized in  \cite{Gimon:2003xk}. One remarkable feature noted in \cite{Gimon:2003xk} is that T-dualities and coordinate twists like (\ref{twist1}), (\ref{twist2}), and (\ref{epemtwist}) do not modify the horizon temperature and its area.  One way to see this is to note that T-dualities acting on a two torus keeps the area in Einstein frame invariant as can be seen from (4.2.23) of \cite{Giveon:1994fu}.  Similarly, coordinate twists simply deforms the complex structure of the torus without changing its area. 

There is however one important subtlety.  We noted that there are three duality chains (\ref{twist1}), (\ref{twist2}), and (\ref{epemtwist}) which leads to the same background (\ref{ttbackground2}) in the zero temperature case.  However,  it is easy to convince oneself that these chains lead to different backgrounds when applied to the finite temperature solution. While it is somewhat cumbersome to carry out this exercise for Maldacena's three charge black hole \cite{Maldacena:1996ky}, one can verify this fact easily by considering a non-extremal fundamental string extended on $x_0$ and $x_1$ and smeared on $y$ direction, and then applying (\ref{twist1}), (\ref{twist2}), and (\ref{epemtwist}). This implies that we have  a large set of finite temperature solutions one can construct by  applying all three twists in some combination.

So at this point, we have discovered a large number of black hole solutions which seems puzzling at first since one expects the black hole solution to be unique once the boundary conditions are fixed. We need to provide a physical interpretation for each of these solutions. Upon closer examination, it appears that although these geometries do approach the zero temperature solution in the large radius region, they behave differently in the subleading asymptotic behavior and can be distinguished by imposing boundary conditions at infinity.  Related discussions on boundary conditions for scalar and vector fields in anti de Sitter space can be found in \cite{Klebanov:1999tb,Marolf:2006nd}. We can in fact see that the behavior of $g_{\mu y}$ and $B_{\mu y}$ fields are playing an important role which is related to the charges and chemical potential of the $U(1)$ global symmetry.

We found the following two constructions to be especially interesting.  For the first case, we begin by parametrizing
\be \epsilon_+ = \epsilon_0 + \epsilon_1, \qquad \epsilon_- = -\epsilon_0 + \epsilon_1  \ . \ee
Now, consider applying (\ref{twist2}) transformation by $(\epsilon_+, \epsilon_-) = (\epsilon_0,-\epsilon_0)$ followed by (\ref{twist1}) by $(\epsilon_+,\epsilon_-) = (\epsilon_1,\epsilon_1)$.
Applying the transformation (\ref{twist1}) involves an explicit shift (\ref{hshift}) in $\lambda$. This is so that the factor in front of $(d \gamma + 2 \epsilon_+ dy)( d \bar \gamma + 2 \epsilon_- dy)$ in (\ref{ttbackground2})  takes the from
\be {1 \over h(\phi)^{-1}-4 \epsilon_+ \epsilon_-} \ . \ee
Since the application of (\ref{twist2}) by $(\epsilon_0,-\epsilon_0)$  already brought this factor into the form
\be {1 \over h(\phi)^{-1}+4 \epsilon_0^2} \ , \ee
the additional shift needed in applying the transformation (\ref{twist1}) is 
\be \lambda \rightarrow \lambda - 4 \epsilon_+ \epsilon_- - 4 \epsilon_0^2  = \lambda - (\epsilon_+  + \epsilon_-)^2 \ , \label{shift1}\ee
instead of (\ref{hshift}). Because we shifted $\lambda$ according to  (\ref{shift1}), the Hagedorn temperature also changes and we find
\be \beta_H \sim  Q_5^{1/2} (\lambda- (\epsilon_+ + \epsilon_-)^2)^{1/2} R\ . \label{ans1} \ee

As a second case, consider first applying (\ref{twist2}) by  $(\epsilon_+, \epsilon_-) = (\epsilon_1,\epsilon_1)$  followed by (\ref{twist1}) by $(\epsilon_+,\epsilon_-) = (\epsilon_0,-\epsilon_0)$ followed by (\ref{twist1}) by $(\epsilon_+,\epsilon_-) = (\epsilon_1,\epsilon_1)$. This time, the shift in $\lambda$ that is needed is 
\be \lambda \rightarrow \lambda - 4 \epsilon_+ \epsilon_- + 4 \epsilon_1^2  = \lambda + (\epsilon_+  - \epsilon_-)^2 \label{shift2}\ee
from which we infer that
\be \beta_H \sim  Q_5^{1/2} (\lambda+ (\epsilon_+ - \epsilon_-)^2)^{1/2} R \ . \label{ans2} \ee

Rather remarkably, we have succeeded in reproducing the $(\lambda,\epsilon_+,\epsilon_-)$ dependence that we found earlier for the fixed charge (\ref{betaA}) and sum over charge (\ref{result1})  ensembles.
This suggests that  (\ref{ans1}) and (\ref{ans2}) correspond the fixed charge and sum over charge ensembles, respectively. We can further justify this identification as follows. In the construction of (\ref{ans1}), the twists and dualities generate mixing between $\gamma_1$ and $y$ coordinates in the form of $g_{1y}$ and $B_{1y}$ which are not directly tied to the energy of the black hole, whereas (\ref{ans2}) generates a mixing between the $\gamma_0$ and $y$ coordinates in the form of $g_{0y}$ and $B_{0y}$ which causes expectation values of the charge to be generated as the energy of the black hole is increased, mimicking the behavior of (\ref{domcharge}).  We should however stress that there is some element of post priori reasoning at work here. We have not systematically analyzed the asymptotic behavior of solutions associated (\ref{ans1}) and (\ref{ans2}) to establish conclusively that they correspond to fixed charge and sum over charge ensembles. At this point, we are merely asserting  the fact that 
explicit procedures exist to construct  holographic backgrounds  with Hagedorn scales  (\ref{ans1}) and (\ref{ans2})  matching the $(\mu,\varepsilon_+, \varepsilon_-)$ dependence of (\ref{betaA}) and (\ref{result1}). 
This fact alone is interesting. It would of course be more interesting to systematically map out the boundary conditions for $g_{\mu y}$,  $B_{\mu y}$, and other fields in this asymptotically linear dilaton background with a twist along the lines of \cite{Klebanov:1999tb,Marolf:2006nd}. The analysis is complicated in part because the twist and the finite temperature effects destroys much of the symmetries to keep the supergravity solutions manageable. Perhaps there is an efficient way to approach this issue, but for now we are leaving this analysis for future work.

Let us make several additional comments before finishing this section.
\begin{enumerate}
\item In  \cite{Chakraborty:2019mdf}, it was observed that in the limit $\lambda - 4 \epsilon_+ \epsilon_- \rightarrow 0$, the geometry (\ref{ttbackground2}) appears to interpolate between $AdS_3$ in the IR to $AdS_2 \times S^1$ when dimensionally reduced along $y$, $T^3$, and $S^3$. From the full ten dimensional perspective, however, it becomes clear that the period of $y$ cycle is going as $1/\sqrt{1-4 h \epsilon_+ \epsilon_-}$   and is becoming large in the large $\phi$ region. Since the period of $y$ cycle is large, one shouldn't dimensionally reduce there. From the oxidized $AdS_3 \times S^1$ perspective, the effect of $\epsilon_\pm$ deformation is merely a twist whose geometric effect is to modify the periodicity conditions. So it seems that the $AdS_2 \times S^1$ geometry does not capture the effective physics of this holographic background.
\item  In  \cite{Chakraborty:2019mdf}, (\ref{Acond}) was interpretable also as the condition for the absence of closed time-like curve. 
 Note, however,  that for $\lambda>0$ and $\epsilon_\pm$ taking real values, (\ref{Acond}) is stronger than the bound $\lambda - 4 \epsilon_+ \epsilon_- >0$ except at the point $\epsilon_+ = \epsilon_-$ where they coincide. 
\item The operation used to construct the supergravity background for the holographic single trace deformed system is identical to the Melvin twist operation which was used in constructing  models known as the Dipole theory which is closely related to non-commutative field theories. In that story, the starting point  was a near horizon limit of a D-brane \cite{Hashimoto:1999ut,Bergman:2000cw,Ganor:2007qh,Song:2011sr}.  Since the construction in this section starts from the NS5 branes and F1 strings, these twists cannot be interpreted exactly as the Dipole field theory, but it is clear that it is in the broad category of non-local field theories related to the Dipole theories via U-duality. 
\item The fact that thermodynamics of Melvin twisted supergravity background is insensitive to the twist follows essentially from the fact that T-dualities and twists acting on a spatial two torus keeps its area in Einstein frame invariant, as was seen in numerous examples \cite{Hashimoto:1999ut,Maldacena:1999mh,Gimon:2003xk,Ganor:2007qh}. Since the area of this torus was directly proportional to the area of the horizon, the thermodynamic relations turned out to be insensitive to the twist. One exception to this pattern of behavior can be found in the T-s-T interpretation of the pure $T \bar T$ deformation discussed in \cite{Apolo:2019zai}. Upon closer look, one sees that the T-s-T transformation discussed  in  section 3.2 of \cite{Apolo:2019zai} involves twisting a torus that involves both spatial and temporal directions. (Strictly speaking,  the authors of \cite{Apolo:2019zai}  twisted and T-dualized along a light like direction, but this can be viewed as a space-like T-duality and a time-like twist that is infinitely boosted.) One can in fact think of the T-s-T of \cite{Apolo:2019zai} as starting with a stack of NS5-F1 and applying the solution generating transformation of \cite{Russo:1996if}. Since the torus being twisted is not entirely proportional to the horizon, there is no reason for the thermodynamics to be unaffected by that twist. 
\end{enumerate}

\section{Discussions}

In this article, we explored various tests and applications of universal formula (\ref{Zdef}) and (\ref{I}) for computing the partition function of $T \bar T$, $J \bar T$, $T \bar J$ deformed conformal field theories.

One main result is the explicit expression (\ref{ZzH}) and (\ref{result1})  for the free energy and the Hagedorn temperature of the deformed compact scalar theory. These quantities were expressible in a compact analytic form by taking the infinite volume limit. These results were shown to be consistent with the expectations from  $x_1 \leftrightarrow (-x_2)$ exchange. We also accounted for the microscopic origin of states with Hagedorn density (\ref{result1}) as arising from the charge sectors (\ref{domcharge}) in the grand canonical ensemble.

We also examined the thermodynamics of single trace $T \bar T$, $J \bar T$, $T \bar J$ deformed $AdS_3 \times S^1 \times T^3 \times S^3$  \cite{Chakraborty:2019mdf} using holographic techniques. We argued that the effects of $J \bar T$ and $T \bar J$ deformation can be realized as a chain of duality and twist transformations starting from the background of pure $T \bar T$ deformed system \cite{Giveon:2017nie}.  In fact, we explained how three different duality chain leads to the same background.  In order to study the thermodynamics, one can construct the finite temperature version of these backgrounds by starting with the finite temperature version of the pure $T \bar T$ deformation \cite{Giveon:2017nie} and applying the same set of twist operations. For the finite temperature background, the three duality chains leads to slightly different backgrounds. We identified specific sequence of dualities and twists which we interpreted as giving rise to the fixed charge and sum over charge ensembles, and found an explicit expression for the Hagedorn temperatures (\ref{ans1}) and (\ref{ans2}) which are in agreement with the dependence on deformation parameters $(\lambda, \epsilon_+, \epsilon_-)$ we found for the free compact boson (\ref{betaA}) and (\ref{result1}).

What we provide in this paper can be thought of as some set of data characterizing the $T \bar T$, $J \bar T$, $T \bar J$ deformed CFT in a handful of examples. It is the case nonetheless that the modular properties provided important consistency checks in carrying out these computations and interpreting the results. One can further attribute the modular properties as being inherited from the sigma model perspective which went into the derivation of (\ref{Zdef}) and (\ref{I}) in \cite{Hashimoto:2019wct}. One can in fact think of the duality and twist operations as moving in $SO(3,3)/SO(3) \times SO(3)$ moduli space of sigma model on $T^2 \times S^1 \times {\cal M}'$.  The $T\bar T$, $J \bar T$, and $T \bar J$ deformations corresponds to a subspace in this moduli space, and it would be interesting to fully map out the $d^2$ dimensional space of $SO(d,d) /SO(d)\times SO(d)$ which arises when there are $n=d-2$ $U(1)$ isometries \cite{Araujo:2018rho}. For $d=3$, the nine parameters appear to correspond to $\lambda$, $\zeta_1$, $\zeta_2$, $\epsilon_+$, $\epsilon_-$, $\chi_1$, $\chi_2$, $r$, and $b$ where $b$ is the NSNS B-field along the $T^2$ and $\chi_1$ and $\chi_2$ are chemical potential for the deformed theory. Some discussion about this chemical potential can be found in Appendix B of \cite{Hashimoto:2019wct}. 
So far, we have been unsuccessful in finding a parameterization of this 9 dimensional space which can make their modular transformation and their physical interpretations simultaneously simple. The perspective based on sigma model and their moduli space should nonetheless be useful for organizing the integrable deformations of these 1+1 dimensional field theories.

\section*{Acknowledgements}
We would like to thank A. Giveon, D. Kutasov and A. Mishra for helpful discussions. The work of SC is supported by the Infosys Endowment for the study of the Quantum Structure of Spacetime. 
AH thanks  IFT-UEPSP for hospitality where part of this work was done.

%\bibliography{hagedorn}\bibliographystyle{utphys}

\begin{thebibliography}{10}

\bibitem{Smirnov:2016lqw}
F.~A. Smirnov and A.~B. Zamolodchikov, ``{On space of integrable quantum field
  theories},'' {\em Nucl. Phys.} {\bf B915} (2017) 363--383,
\href{http://www.arXiv.org/abs/1608.05499}{{\tt 1608.05499}}.
%%CITATION = ARXIV:1608.05499;%%.

\bibitem{Cardy:2018sdv}
J.~Cardy, ``{The $ T\overline{T} $ deformation of quantum field theory as
  random geometry},'' {\em JHEP} {\bf 10} (2018) 186,
\href{http://www.arXiv.org/abs/1801.06895}{{\tt 1801.06895}}.
%%CITATION = ARXIV:1801.06895;%%.

\bibitem{Dubovsky:2018bmo}
S.~Dubovsky, V.~Gorbenko, and G.~Hernández-Chifflet, ``{$ T\overline{T} $
  partition function from topological gravity},'' {\em JHEP} {\bf 09} (2018)
  158,
\href{http://www.arXiv.org/abs/1805.07386}{{\tt 1805.07386}}.
%%CITATION = ARXIV:1805.07386;%%.

\bibitem{Aharony:2018bad}
O.~Aharony, S.~Datta, A.~Giveon, Y.~Jiang, and D.~Kutasov, ``{Modular
  invariance and uniqueness of $T\bar{T}$ deformed CFT},'' {\em JHEP} {\bf 01}
  (2019) 086,
\href{http://www.arXiv.org/abs/1808.02492}{{\tt 1808.02492}}.
%%CITATION = ARXIV:1808.02492;%%.

\bibitem{Datta:2018thy}
S.~Datta and Y.~Jiang, ``{$T\bar{T}$ deformed partition functions},'' {\em
  JHEP} {\bf 08} (2018) 106,
\href{http://www.arXiv.org/abs/1806.07426}{{\tt 1806.07426}}.
%%CITATION = ARXIV:1806.07426;%%.

\bibitem{Guica:2017lia}
M.~Guica, ``{An integrable Lorentz-breaking deformation of two-dimensional
  CFTs},'' {\em SciPost Phys.} {\bf 5} (2018), no.~5, 048,
\href{http://www.arXiv.org/abs/1710.08415}{{\tt 1710.08415}}.
%%CITATION = ARXIV:1710.08415;%%.

\bibitem{Chakraborty:2018vja}
S.~Chakraborty, A.~Giveon, and D.~Kutasov, ``{$ J\overline{T} $ deformed
  CFT$_{2}$ and string theory},'' {\em JHEP} {\bf 10} (2018) 057,
  \href{http://www.arXiv.org/abs/1806.09667}{{\tt 1806.09667}}.

\bibitem{Aharony:2018ics}
O.~Aharony, S.~Datta, A.~Giveon, Y.~Jiang, and D.~Kutasov, ``{Modular
  covariance and uniqueness of $J\bar{T}$ deformed CFTs},'' {\em JHEP} {\bf 01}
  (2019) 085,
\href{http://www.arXiv.org/abs/1808.08978}{{\tt 1808.08978}}.
%%CITATION = ARXIV:1808.08978;%%.

\bibitem{Cavaglia:2016oda}
A.~Cavaglià, S.~Negro, I.~M. Szécsényi, and R.~Tateo, ``{$T
  \bar{T}$-deformed 2D Quantum Field Theories},'' {\em JHEP} {\bf 10} (2016)
  112,
\href{http://www.arXiv.org/abs/1608.05534}{{\tt 1608.05534}}.
%%CITATION = ARXIV:1608.05534;%%.

\bibitem{Bonelli:2018kik}
G.~Bonelli, N.~Doroud, and M.~Zhu, ``{$T \bar{T}$-deformations in closed
  form},'' {\em JHEP} {\bf 06} (2018) 149,
  \href{http://www.arXiv.org/abs/1804.10967}{{\tt 1804.10967}}.

\bibitem{Chakraborty:2019mdf}
S.~Chakraborty, A.~Giveon, and D.~Kutasov, ``{$T\bar{T}$, $J\bar{T}$,
  $T\bar{J}$ and String Theory},'' {\em J. Phys.} {\bf A52} (2019), no.~38,
  384003,
\href{http://www.arXiv.org/abs/1905.00051}{{\tt 1905.00051}}.
%%CITATION = ARXIV:1905.00051;%%.

\bibitem{Giveon:2017nie}
A.~Giveon, N.~Itzhaki, and D.~Kutasov, ``{$ \mathrm{T}\overline{\mathrm{T}} $
  and LST},'' {\em JHEP} {\bf 07} (2017) 122,
\href{http://www.arXiv.org/abs/1701.05576}{{\tt 1701.05576}}.
%%CITATION = ARXIV:1701.05576;%%.

\bibitem{LeFloch:2019rut}
B.~Le~Floch and M.~Mezei, ``{Solving a family of $T\bar{T}$-like theories},''
\href{http://www.arXiv.org/abs/1903.07606}{{\tt 1903.07606}}.
%%CITATION = ARXIV:1903.07606;%%.

\bibitem{Frolov:2019xzi}
S.~Frolov, ``{$T{\overline T}$, $\widetilde JJ$, $JT$ and $\widetilde JT$
  deformations},'' {\em J. Phys. A} {\bf 53} (2020), no.~2, 025401,
  \href{http://www.arXiv.org/abs/1907.12117}{{\tt 1907.12117}}.

\bibitem{Hashimoto:2019wct}
A.~Hashimoto and D.~Kutasov, ``{$ T\overline{T},J\overline{T},T\overline{J} $
  partition sums from string theory},'' {\em JHEP} {\bf 02} (2020) 080,
\href{http://www.arXiv.org/abs/1907.07221}{{\tt 1907.07221}}.
%%CITATION = ARXIV:1907.07221;%%.

\bibitem{Kutasov:1988xb}
D.~Kutasov, ``{Geometry on the Space of Conformal Field Theories and Contact
  Terms},'' {\em Phys. Lett. B} {\bf 220} (1989) 153--158.

\bibitem{Barbon:2020amo}
J.~Barbon and E.~Rabinovici, ``{Remarks on the thermodynamic stability of
  TT-bar deformations},'' \href{http://www.arXiv.org/abs/2004.10138}{{\tt
  2004.10138}}.

\bibitem{Israel:2003ry}
D.~Israel, C.~Kounnas, and M.~P. Petropoulos, ``{Superstrings on NS5
  backgrounds, deformed AdS(3) and holography},'' {\em JHEP} {\bf 10} (2003)
  028, \href{http://www.arXiv.org/abs/hep-th/0306053}{{\tt hep-th/0306053}}.

\bibitem{Apolo:2018qpq}
L.~Apolo and W.~Song, ``{Strings on warped AdS$_{3}$ via $
  \mathrm{T}\bar{\mathrm{J}} $ deformations},'' {\em JHEP} {\bf 10} (2018) 165,
\href{http://www.arXiv.org/abs/1806.10127}{{\tt 1806.10127}}.
%%CITATION = ARXIV:1806.10127;%%.

\bibitem{Apolo:2019yfj}
L.~Apolo and W.~Song, ``{Heating up holography for single-trace $J\bar{T}$
  deformations},'' {\em JHEP} {\bf 01} (2020) 141,
  \href{http://www.arXiv.org/abs/1907.03745}{{\tt 1907.03745}}.

\bibitem{Maldacena:1996ky}
J.~M. Maldacena, {\em {Black holes in string theory}}.
\newblock PhD thesis, Princeton U., 1996.
\newblock
\href{http://www.arXiv.org/abs/hep-th/9607235}{{\tt hep-th/9607235}}.
\newblock
%%CITATION = HEP-TH/9607235;%%.

\bibitem{Apolo:2019zai}
L.~Apolo, S.~Detournay, and W.~Song, ``{TsT, $T\bar{T}$ and black strings},''
  \href{http://www.arXiv.org/abs/1911.12359}{{\tt 1911.12359}}.

\bibitem{Gimon:2003xk}
E.~G. Gimon, A.~Hashimoto, V.~E. Hubeny, O.~Lunin, and M.~Rangamani, ``{Black
  strings in asymptotically plane wave geometries},'' {\em JHEP} {\bf 08}
  (2003) 035,
\href{http://www.arXiv.org/abs/hep-th/0306131}{{\tt hep-th/0306131}}.
%%CITATION = HEP-TH/0306131;%%.

\bibitem{Giveon:1994fu}
A.~Giveon, M.~Porrati, and E.~Rabinovici, ``{Target space duality in string
  theory},'' {\em Phys. Rept.} {\bf 244} (1994) 77--202,
  \href{http://www.arXiv.org/abs/hep-th/9401139}{{\tt hep-th/9401139}}.

\bibitem{Klebanov:1999tb}
I.~R. Klebanov and E.~Witten, ``{AdS / CFT correspondence and symmetry
  breaking},'' {\em Nucl. Phys. B} {\bf 556} (1999) 89--114,
  \href{http://www.arXiv.org/abs/hep-th/9905104}{{\tt hep-th/9905104}}.

\bibitem{Marolf:2006nd}
D.~Marolf and S.~F. Ross, ``{Boundary Conditions and New Dualities: Vector
  Fields in AdS/CFT},'' {\em JHEP} {\bf 11} (2006) 085,
  \href{http://www.arXiv.org/abs/hep-th/0606113}{{\tt hep-th/0606113}}.

\bibitem{Hashimoto:1999ut}
A.~Hashimoto and N.~Itzhaki, ``{Noncommutative Yang-Mills and the AdS / CFT
  correspondence},'' {\em Phys. Lett.} {\bf B465} (1999) 142--147,
\href{http://www.arXiv.org/abs/hep-th/9907166}{{\tt hep-th/9907166}}.
%%CITATION = HEP-TH/9907166;%%.

\bibitem{Bergman:2000cw}
A.~Bergman and O.~J. Ganor, ``{Dipoles, twists and noncommutative gauge
  theory},'' {\em JHEP} {\bf 10} (2000) 018,
\href{http://www.arXiv.org/abs/hep-th/0008030}{{\tt hep-th/0008030}}.
%%CITATION = HEP-TH/0008030;%%.

\bibitem{Ganor:2007qh}
O.~J. Ganor, A.~Hashimoto, S.~Jue, B.~S. Kim, and A.~Ndirango, ``{Aspects of
  Puff Field Theory},'' {\em JHEP} {\bf 08} (2007) 035,
  \href{http://www.arXiv.org/abs/hep-th/0702030}{{\tt hep-th/0702030}}.

\bibitem{Song:2011sr}
W.~Song and A.~Strominger, ``{Warped AdS3/Dipole-CFT Duality},'' {\em JHEP}
  {\bf 05} (2012) 120, \href{http://www.arXiv.org/abs/1109.0544}{{\tt
  1109.0544}}.

\bibitem{Maldacena:1999mh}
J.~M. Maldacena and J.~G. Russo, ``{Large N limit of noncommutative gauge
  theories},'' {\em JHEP} {\bf 09} (1999) 025,
  \href{http://www.arXiv.org/abs/hep-th/9908134}{{\tt hep-th/9908134}}.

\bibitem{Russo:1996if}
J.~Russo and A.~A. Tseytlin, ``{Waves, boosted branes and BPS states in m
  theory},'' {\em Nucl. Phys. B} {\bf 490} (1997) 121--144,
  \href{http://www.arXiv.org/abs/hep-th/9611047}{{\tt hep-th/9611047}}.

\bibitem{Araujo:2018rho}
T.~Araujo, E.~Colgáin, Y.~Sakatani, M.~Sheikh-Jabbari, and H.~Yavartanoo,
  ``{Holographic integration of $T \bar{T}$ \& $J \bar{T}$ via $O(d,d)$},''
  {\em JHEP} {\bf 03} (2019) 168,
  \href{http://www.arXiv.org/abs/1811.03050}{{\tt 1811.03050}}.

\end{thebibliography}

\providecommand{\href}[2]{#2}\begingroup\raggedright\endgroup

\end{document}